\documentclass[12pt,a4paper,twoside]{article}

\usepackage{publication.twoside} 

\begin{document}

~\vspace{0.5cm}
\begin{center}
\Large{\sc Parameter estimation of quantum processes using convex optimization}
\vspace{0.5cm}\\ 
\large{\textit{Research report}}
\vspace{0.5cm}\\ 
\textbf{\normalsize{Gábor Balló}}\\
\small{Department of Electrical Engineering and Information Systems,\\University of Pannonia,\\Veszprém, Hungary}
\vspace{0.5cm}\\ 
\textbf{\normalsize{Katalin M. Hangos}}\\
\small{Process Control Research Group,\\Computer and Automation Research Institute,\\Budapest, Hungary}
\vspace{0.5cm}\\ 
\large{\today}
\thispagestyle{empty} 
\end{center}
\vspace{1cm}
\begin{abstract}
A convex optimization based method is proposed for quantum process tomography, in the case of known channel model structure, but unknown channel parameters. The main idea is to select an affine parametrization of the Choi matrix as a set of optimization variables, and formulate a semidefinite programming problem with a least squares objective function. Possible convex relations between the optimization variables are also taken into account to improve the estimation.

Simulation case studies show, that the proposed method can significantly increase the accuracy of the parameter estimation, if the channel model structure is known. Beside the convex part, the determination of the channel parameters from the optimization variables is a nonconvex step in general. In the case of Pauli channels however, the method reduces to a purely convex optimization problem, allowing to obtain a globally optimal solution. 
\end{abstract}
\newpage 
\thispagestyle{empty}
~
\newpage

\markboth{\MakeUppercase{\leftmark}}{} 
\tableofcontents  

\newpage
\section{Introduction}
\label{sec:intro}
In these days, the building of quantum computers that can be used to solve realistic, large scale problems---including the breaking of cryptographic codes and the simulation of complex quantum systems---has two main difficulties from the theoretical point of view. The first is decoherence, in other words the unavoidable coupling of quantum systems with their environment, which leads to the altering or even the complete destruction of the quantum system's state, thus causing errors in the calculation. The other main obstacle is the lack of ability to fully manipulate and extract full information from the physical system, i.e. the control and estimation of the quantum states and channels, called quantum processes. This latter problem has to be handled by the formalism of the modern methods of system- and control theory. It is certainly not a trivial problem, because in quantum mechanics, one has to face the difficulty of the threatment of measurements. Namely, that no measurement can be carried out on a quantum system without disturbing the state of the system itself.  

The task of the estimation of quantum channels---commonly known as quantum process tomography (QPT) \cite{Nielsen-book}---got a significant attention over about the last ten years. It is undoubtedly a fundamental problem of quantum information theory, as it has considerable relevance not only in quantum computers, but also in the field of quantum communication and cryptography. For example, quantum communication channels usually rely on a priori knowledge of the channel properties. 
 
The problem of quantum process tomography was investigated by several authors \cite{Nielsen-book,Dariano-2003,Paris-book}. The work \cite{Mohseni-2008} gives a comprehensive survey on the different strategies used for process tomography (or channel estimation). The problem can essentially be formulated in two type of methods: direct, and indirect. In the indirect method, we trace the problem back to quantum state tomography, i.e. the information about the unknown quantum channel is obtained by sending known probe quantum systems through the channel, and performing state tomography on the output states. In contrast, in the direct method, the experiments directly give information about the channel, without the need for a state tomography step.

From a methodological point of view, there are two principally different approaches to the problem of quantum tomography, the \emph{statistical approach} and the convex optimization based approach \cite{Dariano-2003}. The former gives information on the statistics of the estimate and on its covariance matrix, but it has the drawback, that it is hard to compute in higher dimensions. In spite of this, majority of the existing methods belong to this category. 

In contrast to this, an \emph{optimization based method} does not give as much information, but it is relatively easy to compute. This approach has been pursued in the work \cite{Sacchi-2001} where the problem of channel estimation (in the form of the Choi matrix) is considered, assuming a completely general channel, thus without any assumption on the inner structure of the Choi matrix. The author uses random input states, and random measurements on the output, and formulates a maximum likelihood problem. A similar method is used in the work \cite{Kosut-2004}, which formulates the task of process tomography as a least squares problem, which is convex. It also uses the Choi matrix as optimization variable, thus searches the optimal channel in the convex set of all CPTP maps using multiple input-measurement pairs. 

However, as it is stated in \cite{Sasaki-2002}, it is a reasonable assumption to consider only a certain family of channels given with a model, based on a priori knowledge about the structure of the channel. The aim of this work is to develop a method which is capable of incorporating these constraints into the channel tomography problem, while still remaining -- at least partially in the general case -- solvable by convex optimization. 

The accuracy of the estimation is measured with the empirical covariance matrix. 

The rest of this paper is organised as follows. In Section \ref{sec:qpt.as.opt.problem} the formal description of quantum process tomography is given, and a least squares objective is derived for the solution. In Section \ref{sec:est.model.families} the variables and constraints of the convex optimization problem are considered, and the extraction of the channel parameters is discussed. In Section \ref{sec:case.studies} the procedure is demonstrated on several case studies. Finally, in Section \ref{sec:conclusion} the conclusions are given, and the direction of further work is discussed.


\section[QPT as an optimization problem]{Quantum process tomography as an optimization\\ problem}
\label{sec:qpt.as.opt.problem}
In this work, the indirect procedure of process tomography is followed, mostly relying on the work \cite{Kosut-2004}. The formal mathematical description of process tomography, i.e. the so called \emph{tomography configuration} contains the following elements: 
\begin{itemize}
\item A known input density operator $\rho$ on the Hilbert space $\spa{H}$ of the system. 
\item The unknown quantum channel $\cha{E}: \spa{L}(\spa{H})\rightarrow\spa{L}(\spa{H})$, which is to be estimated. The channel can be written for example in the Kraus representation, in which case the output of $\cha{E}$ is $\sigma=\cha{E}(\rho)=\sum_i\op{E}_{i}\rho\op{E}_{i}^\dagger$, where the $\op{E}_{i}$ operators are the \emph{operator elements} of the channel $\cha{E}$. These must satisfy the trace preserving constraint $\sum_i\op{E}_i^\dagger\op{E}_i=\hop{1}$, where $\hop{1}$ is the identity operator.\footnote{Here $^\dagger$ denotes the adjoint of the operator.}
\item A set of POVMs. A POVM $\textbf{M}=\{\op{M}_{\alpha}\}$ is a set of positive operators, with which we can perform quantum measurement on the channel output state $\sigma$. In order to be able to uniquely identify the channel output state, an important requirement in quantum process tomography is that the measurements must be \emph{tomographically complete}. This means that the measured POVMs must provide all the information about the output state, and thus on the channel. Such a set of measurement operators is sometimes called a \emph{quorum}.
\end{itemize}
Note that we can use multiple different tomography configurations, i.e. different input states and POVMs in order to achieve better estimation on $\cha{E}$. In this work, the input-POVM pair corresponding to the $\gamma$th configuration is denoted by $\rho_\gamma$ and $\textbf{M}_\gamma$. 


\subsection{Data collection}
\label{subsec:data.collection}
\begin{figure}
\im[width=\textwidth]{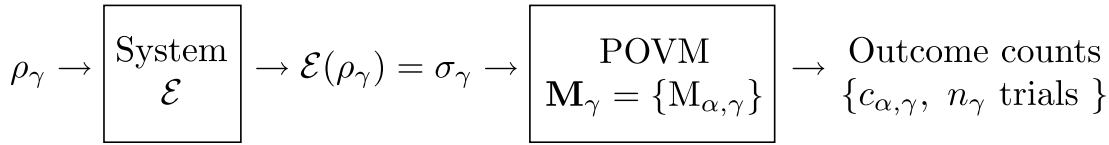}
\caption{The scheme of data collection for process tomography.\label{fig:data.collection}}
\end{figure}

The first stage of process tomography is the collection of the measurement data into a measurement record. The measurements are performed in each $\gamma$ configuration $n_\gamma$ times independently. This scheme can be seen in Figure \ref{fig:data.collection}. 

During data collection, the different $\alpha$ outcomes of the measurements in the configuration $\gamma$ are counted in the variable $c_{\alpha,\gamma}$, and put in the measurement record $\textbf{D}$. Then obviously 
\[  
\sum_\alpha c_{\alpha,\gamma} = n_\gamma\ .
\]
Thus, we have to perform a total number of $n_\tn{tot}=\sum_\gamma n_\gamma$ independent measurements. The estimator $\hat{\cha{E}}$ of the channel $\cha{E}$ will be calculated from these measurement outcomes, based on some estimation procedure.


\subsection{Least Squares estimation}
\label{subsec:ls.estimation}
The next step of the tomography problem is to choose a suitable estimation procedure. The least squares is a popular method, because it is easy to implement, and it results in an estimation procedure, which can be used in higher dimensions, and for multiple parameters. So in this work it is used for quantum process estimation. The exact form of the LS objective function for the process tomography problem is derived in the following. 

If we write the $\sigma_{\gamma}=\cha{E}(\rho_\gamma)$ output of the channel in Kraus representation:
\[
\sigma_{\gamma}=\sum_i \op{E}_{i}\rho_\gamma\op{E}_{i}^\dagger\ ,
\]
then the probability density function of the measurement results can be modeled by  
\begin{equation}
\label{eq:tomo.measurement.prob}
p_{\alpha,\gamma}=\trace{\big(\sigma_{\gamma}\op{M}_{\alpha,\gamma}\big)}=\trace{\left(\sum_i \op{E}_{i}^\dagger\op{M}_{\alpha,\gamma}\op{E}_{i}\rho_\gamma\right)}\ .
\end{equation}

In this expression, the optimization variables would be the Kraus operator elements $\op{E}_i$. However, as these do not form a convex set, the resulting LS optimization problem would be nonconvex. To overcome this difficulty, we can choose the Choi matrix as optimization variable. 

A common definition of the Choi matrix is through the use of the Jamio{\l}kowsky isomorphism \cite{Jamiolkowski-1972}. Let $\spa{H}_1$ and $\spa{H}_2$ be Hilbert spaces, and let $\op{A}$ be a bounded linear operator in $\spa{L}(\spa{H}_1,\spa{H}_2)$. Then we can associate $\op{A}$ with a vector in the Hilbert-space $\spa{H}_1\otimes\spa{H}_2$:
\[
\kett{\op{A}}\coloneqq\sum_{i}\op{A}\ket{i_1}\ket{i_2}\ ,
\]
where $\{\ket{i_1}\}$ and $\{\ket{i_2}\}$ are the respective bases of the spaces $\spa{H}_1$ and \spa{H}$_2$.\footnote{The \kett{.} notation indicates that these vectors represent operators.} By this, noting that in our case $\spa{H}_1=\spa{H}_2=\spa{H}$ is the Hilbert space of the input state, the Choi matrix of the channel $\cha{E}$ will be 
\[
\op{X}_\cha{E}\coloneqq\sum_k\kettbra{\op{E}_k}{\op{E}_k}\ ,
\]
where the vector $\kett{\op{E}_k}$ in $\spa{H}\otimes\spa{H}$ is associated to the operator $\op{E}_k$.
The condition of complete positivity can then be written as
\[
\op{X}_\cha{E}\geq 0\ ,
\]
and the channel is trace preserving if 
\[
\trace_{\spa{H}_{(2)}}(\op{X}_\cha{E})=\hop{1}\in\spa{L}(\spa{H}_{(1)})\ .
\]

The scalar product of these ``double-ket'' vectors is defined in a natural way:
\begin{equation}
\label{eq:Hilbert--Schmidt.scalar.prod.def}
\brakett{\op{A}}{\op{B}}\coloneqq\trace\big(\op{A}^\dagger\op{B}\big)\ , 
\end{equation}
what is by definition the Hilbert--Schmidt scalar product of the two operators. We can also identify a useful rule easily\footnote{Here $^\tp$ denotes the transposition.}: 
\begin{equation}
\label{eq:choi.rule.1}
\big(\op{A}\otimes\op{B}\big)\kett{\op{C}}=\kett{\op{B}\op{C}\op{A}^\tp}\ ,
\end{equation}

From the above we see that an isomorphism can be made between the density operators on $\spa{L}(\spa{H})$ and the pure states of the space $\spa{H}\otimes\spa{H}$.

Now, using the above, we can continue the derivation of \eqref{eq:tomo.measurement.prob}. By the use of relations \eqref{eq:Hilbert--Schmidt.scalar.prod.def} and \eqref{eq:choi.rule.1} we get
\[
\begin{split}
p_{\alpha,\gamma}&=\sum_i\trace{\left(\op{E}_{i}^\dagger\op{M}_{\alpha,\gamma}\op{E}_{i}\rho_\gamma\right)}=\\
&=\sum_i\brakett{\op{M}_{\alpha,\gamma}^\dagger\op{E}_{i}}{\op{E}_{i}\rho_{\gamma}}=\\
&=\sum_i\braa{\op{E}_{i}}(\hop{1}\otimes\op{M}_{\alpha,\gamma}^\dagger)(\rho_\gamma^\tp\otimes\hop{1})\kett{\op{E}_{i}}=\\
&=\sum_i\trace\left[\underbrace{(\rho_\gamma^\tp\otimes\op{M}_{\alpha,\gamma}^\dagger)}_{\op{R}_{\alpha,\gamma}}\kettbra{\op{E}_{i}}{\op{E}_{i}}\right]=\\
&=\trace(\op{R}_{\alpha,\gamma}\op{X}_{\cha{E}})\ ,
\end{split}
\]
where $\op{X}_{\cha{E}}$ is the Choi matrix of the channel $\cha{E}$, and the operator $\op{R}_{\alpha,\gamma}$ depends on the channel input $\rho$ and on the measured POVM elements in configuration $\gamma$. 

Proceeding to the derivation of the LS objective function, a statistics for estimating the probability $p_{\alpha,\gamma}$ can be given by the 
\begin{equation}
\label{eq:ls.relative.frequency}
\hat{p}_{\gamma}(\alpha)=\frac{c_{\alpha,\gamma}}{n_\gamma}
\end{equation} 
relative frequency that can be calculated from the measurement results. The variance of this unbiased estimate after $n_\gamma$ independent measurements is known to be 
\[
\mathrm{Var}\big(\hat{p}_{\gamma}(\alpha)\big)=\frac{1}{n_\gamma}p_{\alpha,\gamma}\big(1-p_{\alpha,\gamma}\big)\ ,
\]
because $\hat{p}_\gamma(\alpha)$ has a binomial distribution. These show that for large $n_\gamma$, $\hat{p}_{\gamma}(\alpha)\rightarrow p_{\alpha,\gamma}$ and $\mathrm{Var}\big(\hat{p}_{\gamma}(\alpha)\big)$ tends to $0$ as $n_\gamma\rightarrow\infty$, so $\hat{p}_{\gamma}(\alpha)$ is a reasonable unbiased estimate of the real value $p_{\alpha,\gamma}$. This leads to formulating the parameter estimation problem as the following least squares objective: 
\begin{gather}
\label{eq:ls.objective}
\arg\min_{\op{X}_{\cha{E}}} V_{\tn{LS}}(\op{X}_{\cha{E}})=\sum_{\alpha,\gamma}\trace\big(\op{P}^\tn{emp}_{\alpha,\gamma}-\op{R}_{\alpha,\gamma}\op{X}_{\cha{E}}\big)^2\ ,\\
\tn{so that}\quad\op{X}_\cha{E}\geq 0,\quad\trace_{\spa{H}}(\op{X}_\cha{E})=\hop{1}\ \nonumber
\end{gather}
where the diagonal matrix $\op{P}^\tn{emp}_{\alpha,\gamma}=\hat{p}_{\gamma}(\alpha)\frac{\hop{1}_{\spa{H}\otimes\spa{H}}}{\dim(\spa{H}\otimes\spa{H})}$ is defined for simplification purpose.

This problem is a convex optimization problem in the Choi matrix $\op{X}_\cha{E}$, thus it can be solved relatively easily using existing numerical algorithms \cite{Vandenberghe-1996,Audenaert-2002}. 


\section{Estimation of channel model families}
\label{sec:est.model.families}

It is reasonable to assume that we know a model type of the channel, and only the unknown values of the parameters of this model have to be estimated. In such a problem, the above derived least squares objective cannot be used directly, as it assumes a completely general channel model, and estimates the elements of the Choi matrix. Thus, if the task is to estimate some specific model parameters, this method can suffer significantly from overparametrization. 

As a possible solution, we can study the internal structure of the Choi matrix, and use this information to select more appropriate, model specific parameters for optimization. Effectively, this should reduce the set of optimal solutions of problem \eqref{eq:ls.objective} to solutions, which are consistent with the desired model family. 


\subsection{The complex LMI constraint}
\label{ssec:affine}
The natural choice would be to select just the unknown channel parameters, however it can be easily seen, that this choice would ruin convexity, as the Choi matrix can be an arbitrarily nonconvex function of these in the most general case. 

Thus, instead of this, the following method is used. Let $h_1(\theta),\dots,h_m(\theta)$ denote functions of the channel parameters, and let $H_0,H_1,\dots,H_m$ denote constant Hermitian matrices. Then we can expand the Choi matrix as an affine function:
\begin{equation}
\label{eq:complex.lmi}
\op{X}_\cha{E}=\sum_kH_kh_k(\theta)+H_0
\end{equation}
It can be seen, that if we consider the functions $h_k(\theta)$ as optimization variables, we can replace the completely unknown matrix variable $\op{X}_\cha{E}$ in problem \eqref{eq:ls.objective} with a matrix, which has a known structure up to an affine approximation. 

Thus, the positivity constraint in \eqref{eq:ls.objective} turns into a complex LMI constraint, and the trace preserving constraint can in turn be omitted, as if an exact channel model is known, it can always be taken into account in the model construction.   


\subsection{Convex constraints}
It can occur, that some of the functions $h_k(\theta)$ depend on one or more other functions. If this relation $f$ is convex, then we can use it to define additional constraints to make the estimation more accurate. We can determine a minimal set $\mathcal{V}=\{h_l(\theta)\}$, where $l\in I_\mathcal{V}$. The set $I_\mathcal{V}$ contains the indexes of those functions among $h_k(\theta)$ ($k=1,\dots,m$), which do not depend on each other through a convex relation, thus are independent in this sense. These can in general be written as
\[
h_k(\theta)=f^{\tn{cvx}}\big(h_1(\theta),\dots,h_{m'}(\theta)\big),\quad h_l(\theta)\in\mathcal{V},\quad l=1,\dots,m'
\]
Thus, if we have convex relations between the functions $h_k(\theta)$ chosen as optimization variables, then the identification of these relations can help to reduce the number of these variables, which can be helpful in the parameter extraction step (see subsection \ref{ssec:param.extract}.).

However, these relations cannot be added simply as equations to the problem \eqref{eq:ls.objective}, because a strictly convex function with equality is not a convex constraint. In these cases we can relax the equality, and add the convex constraint to the problem as inequality. 

If $n_c$ such constraints are found, then we obtain the following optimization problem (omitting the trace preserving constraint):
\begin{gather}
\arg\min_{\n{h}(\theta)} V_{\tn{LS}}(\op{X}_{\cha{E}})\ \nonumber,\\
\tn{so that}\quad\op{X}_{\cha{E}}\geq 0\ ,\label{eq:ls.objective.convex}\\ 
\tn{and}\quad f_k^{\tn{cvx}}\big(h_1(\theta),\dots,h_{m'}(\theta)\big)\leq h_k(\theta),\quad k=1,\dots,n_c\ \nonumber
\end{gather}
After solving this problem we get an optimal value $v_0^*=V_{\tn{LS}}(\op{X}^*_{\cha{E}})$ for the objective, where the $^*$ denotes the optimality. Then we formulate for each $k$ the following auxiliary problem:
\begin{gather}
\arg\min_{\n{h}(\theta)} h_{k}(\theta)\ \nonumber,\\
\tn{so that}\quad\op{X}_\cha{E}\geq 0\ ,\label{eq:aux.convex}\\
\tn{and}\quad f_k^{\tn{cvx}}\big(h_1(\theta),\dots,h_{m'}(\theta)\big)\leq h_k(\theta),\quad k=1,\dots,n_c\ \nonumber,\\
\tn{and}\quad v_k=V_{\tn{LS}}(\op{X}_{\cha{E}})\leq v_{k-1}^* \nonumber
\end{gather}
By solving these for each $k$ in succession, we make sure that the additional convex constraints in \eqref{eq:ls.objective.convex} get as close to equality as possible, thus making the parameter estimation procedure more accurate. However it would require further studies to see whether this method always guarantees a global optimum in the original objective, and whether it is reachable in finite steps. In practice, the effectiveness of this heuristics is based only on the fact, that in case of the optimal solution (in the model of the real channel), all of the convex constraints are saturated. 


\subsection{Determining the model parameters}
\label{ssec:param.extract}
After we performed all the needed optimization steps and obtained an optimal Choi matrix $\op{X}^*_\cha{E}$ together with the optimal variables $h_k^*$, the next step of parameter estimation is to fit $h_k^*$ to the predefined optimization variables $h_k(\theta)$. In essence, this problem can be written as  
\begin{equation*}
\label{eq:paramextract}
\arg\min_{\theta}\Vert\n{h}^*-\n{h}(\theta)\Vert\ ,
\end{equation*}
where the norm is arbitrary.

Of course, only those parameters has to be used here which are in the set $\mathcal{V}$ of independent variables. Using the Euclidean norm, this problem thus can be reduced to the following. Let $I_\mathcal{V}$ denote the index set of the independent optimization variables: 
\begin{equation}
\label{eq:paramextract.hs}
\arg\min_{\theta} \sum_{l\in I_\mathcal{V}} \big(h_l^*-h_l(\theta)\big)^2\ 
\end{equation}
This is a nonlinear least squares problem, which is not even convex in the general case, because of the $h_l(\theta)$ functions. If however these functions are convex, then the problem \eqref{eq:paramextract.hs} will be convex too. 

Note that the number of independent optimization variables determines the complexity of this nonlinear least squares problem, as having fewer variables result in fewer nonlinear equations.  


\section{Case studies}
\label{sec:case.studies}
The aim of the simulation experiments was to analyze the effect of additional constraints on performance of the numerical optimization based estimation of quantum channels. Results were generated in MATLAB environment, using simulated random measurement data. The optimization problem \eqref{eq:ls.objective.convex} and \eqref{eq:aux.convex} were solved using YALMIP modeling language \cite{YALMIP} and the SDPT3 solver \cite{SDPT3}. The extraction of the model parameters, i.e. the solving of \eqref{eq:paramextract.hs} was in general done by the built in nonlinear least squares solver of MATLAB, solving it multiple times using random initial conditions to achieve a global minimum with high probability. 

\subsection{Tomography configurations}
The experiments were set up as follows.
\begin{itemize}
\item The used input states were all pure states. 
\item To obtain a tomographically complete measurement, appropriate POVMs were selected on the Hilbert space of the system, and were used for measurement. In two dimensions, these are the \emph{Pauli matrices}, and observables with similar properties in higher dimensions. Each of these can be decomposed into a POVM, which can be used in one configuration. Note that the tomographically complete measurement in this setting may not require an operator basis of observables to be used. The knowledge about the channel structure may allow us to identify parts of the state space where the effect of the channel is the same, thus requiring only less observables (see case study in subsection \ref{subsec:gen.pauli}). The only important thing is to get information on all of the channel parameters. 
\item The total $n_\tn{tot}$ number of measurements was distributed among all the configurations equally, i.e. for each configuration $\gamma$, an equal number of experiment were used. 
\end{itemize}
Each experiment setup was repeated five times and their average was taken. Each of the estimated process Choi matrices $\op{X}_\cha{E}$ and channel parameters were analyzed using the following four estimation performance measuring quantities:
\begin{itemize}
\item Fidelity of the estimated and ideal channel output: $F\big(\hat{\cha{E}}(\rho),\cha{E}(\rho)\big)$. The fidelity of two quantum states is defined by 
\[
F(\rho,\sigma)=\trace\sqrt{\rho^\frac{1}{2}\sigma\rho^\frac{1}{2}}\ .
\]
\item The empirical mean $\bar{\theta}$ of the estimated parameters $\hat{\theta}$. 
\item The empirical covariance matrix of the estimated parameters $\hat{\theta}$. This could be used only in cases, where the empirical covariance could be made diagonal with appropriately selected measurement observables. The reason for this is that in other cases, the estimated parameters were correlated. One way to obtain a diagonal empirical covariance is, that the output state parametrization has to be compatible with the channel parametrization in the sense that each state parameter is affected only by one channel parameter. In this case we can use that set of operators for measurement in which the state parametrization is given, and through this compatibility, we will be able to estimate the channel parameters independently \cite{Petz-2008}. 

Thus, only the diagonal elements of the covariance matrix, i.e. the variances of the parameters were computed:
\[
\tn{Var}(\hat{\theta}_i)=\frac{1}{4}\sum_{j=1}^5(\hat{\theta}_{i,j}-\bar{\theta}_{i})^2\ .
\]
\item The Hilbert--Schmidt norm of the estimation error: $\Vert\hat{\op{X}}-\op{X}\Vert$.
\end{itemize}


\subsection{The generalized amplitude damping channel}
\label{ssec:gen.amp}
The Choi matrix of the generalized amplitude damping channel is
\[ 
\op{X}_{\cha{E}}=\left[ \begin {array}{cccc} 1-\gamma+p\,\gamma&0&0&\sqrt {1-\gamma}\\\noalign{\medskip}0&\gamma-p\,\gamma&0&0\\\noalign{\medskip}0&0&p\,\gamma&0\\\noalign{\medskip}\sqrt {1-\gamma}&0&0&-p\,\gamma+1\end {array} \right]\ .
\] 
Here the parameter $\gamma$ gives the strength of the noise effect in the interval $[0,1]$, and $p$ parameterizes the output state for $\gamma=1$ also in the interval $[0,1]$. This is $\ket{0}$ when $p=1$, and $\ket{1}$ when $p=0$ in the computational basis. 

Using \eqref{eq:complex.lmi}, the above Choi matrix can be decomposed in the following way:
\begin{align*}
&H_0=
{\scriptsize 
\left[ 
\begin {array}{cccc} 
 1 &  0  &  0  &  0\\ \noalign{\medskip}
 0 &  0  &  0  &  0\\ \noalign{\medskip}
 0 &  0  &  0  &  0\\ \noalign{\medskip}
 0 &  0  &  0  &  1
\end{array}
\right]
},
&&H_1=
{\scriptsize 
\left[ 
\begin {array}{cccc} 
-1 &  0  &  0  &  0\\ \noalign{\medskip}
 0 &  1  &  0  &  0\\ \noalign{\medskip}
 0 &  0  &  0  &  0\\ \noalign{\medskip}
 0 &  0  &  0  &  0
\end{array}
\right]
},\\
&H_2=
{\scriptsize 
\left[ 
\begin {array}{cccc} 
 1 &  0  &  0  &  0\\ \noalign{\medskip}
 0 & -1  &  0  &  0\\ \noalign{\medskip}
 0 &  0  &  1  &  0\\ \noalign{\medskip}
 0 &  0  &  0  & -1
\end{array}
\right]
},
&&H_3=
{\scriptsize 
\left[ 
\begin {array}{cccc} 
 0 &  0  &  0  &  1\\ \noalign{\medskip}
 0 &  0  &  0  &  0\\ \noalign{\medskip}
 0 &  0  &  0  &  0\\ \noalign{\medskip}
 1 &  0  &  0  &  0
\end{array}
\right]
},
\end{align*}
and the optimization variables will be
\[
h_1=\gamma,\quad h_2=p\gamma,\quad h_3=\sqrt{1-\gamma}\ .
\]

Notice that we can find a convex relation between these variables, thus we can further reduce their number: 
\[
h_3^2-1=-h_1
\]

If we now put these into the problem \eqref{eq:ls.objective.convex}, we get the following optimization problem:
\begin{gather}
\arg\min_{\n{h}(\theta)} v_0=V_{\tn{LS}}(\op{X}_\cha{E})\ \nonumber,\\
\tn{so that}\quad\op{X}_{\cha{E}}\geq 0,\label{eq:ls.objective_amp}\\
\tn{and}\quad h_3^2-1\leq-h_1\ \nonumber
\end{gather}

The type \eqref{eq:aux.convex} auxiliary problem will be:
\begin{gather*}
\arg\min_{\n{h}(\theta)}\,-h_1\ \nonumber,\\
\tn{so that}\quad\op{X}_\cha{E}\geq 0,\label{eq:aux.amp}\\
\tn{and}\quad h_3^2-1\leq-h_1\ \nonumber,\\
\tn{and}\quad V_{\tn{LS}}(\op{X}_{\cha{E}})\leq v_0^*\nonumber
\end{gather*}

After solving these, the final step is the parameter extraction using \eqref{eq:paramextract.hs}. As the independent set $\mathcal{V}=\{h_1,h_2\}$ of optimization variables contains nonconvex functions, this problem will also be nonconvex. However, in this simple two dimensional case, we can see that the solution of \eqref{eq:paramextract.hs} will be explicitly
\[
\gamma=h_1,\quad p=\frac{h_2}{h_1}\ .
\]

The input state used in this example was the pure state with the Bloch vector $\frac{1}{\sqrt{3}}[1,1,1]^\tp$. The applied observables were the Pauli matrices
\[
\sigma_1=\left[ \begin {array}{cc} 0&1\\\noalign{\medskip}1&0\end {array} \right], \quad \sigma_2=\left[ \begin {array}{cc} 0&-\ii\\\noalign{\medskip}\ii&0\end {array} \right], \quad  \sigma_3=\left[ \begin {array}{cc} 1&0\\\noalign{\medskip}0&-1\end {array} \right]\ .
\]  
These form a tomographically complete set of observables for the two dimensional case. Each was used as a two-element POVM in a different configuration, thus a total of three configurations were used. Unfortunately, this set does not allow the indepencent estimation of the parameters for this channel, thus the empirical covariance matrix cannot be used in this case as performance indicator. 

The resulting estimations, and the used characteristic quantities are plotted for three different pairs of exact parameter values in Figures \ref{fig:amp1}-\ref{fig:amp3}. The chosen $(\gamma,p)$ pairs were $(0.7,0.3)$, $(0,1)$ and $(1,0)$. These values were selected to represent both the interior and border of the parameter space. 
\begin{figure}
\begin{center}
\subfloat[]
{   
    \im{amp1-fid.eps}
    \label{fig:amp1-fid} 
}
\subfloat[]
{   
    \im{amp1-mea.eps}
    \label{fig:amp1-mea}
}\\
\subfloat[]
{   
    \im{amp1-var.eps}
    \label{fig:amp1-var}
}
\subfloat[]
{   
    \im{amp1-hil.eps}
    \label{fig:amp1-hil}
}
\caption{The parameters are $\gamma=0.7$ and $p=0.3$. \subref{fig:amp1-fid} shows the fidelity of the estimated and ideal channel output, \subref{fig:amp1-mea} shows the expected value of the estimated parameters, \subref{fig:amp1-var} shows the variance of the estimated parameters, and \subref{fig:amp1-hil} shows the Hilbert--Schmidt norm of the estimation error.\label{fig:amp1}}
\end{center}
\end{figure}

\begin{figure}
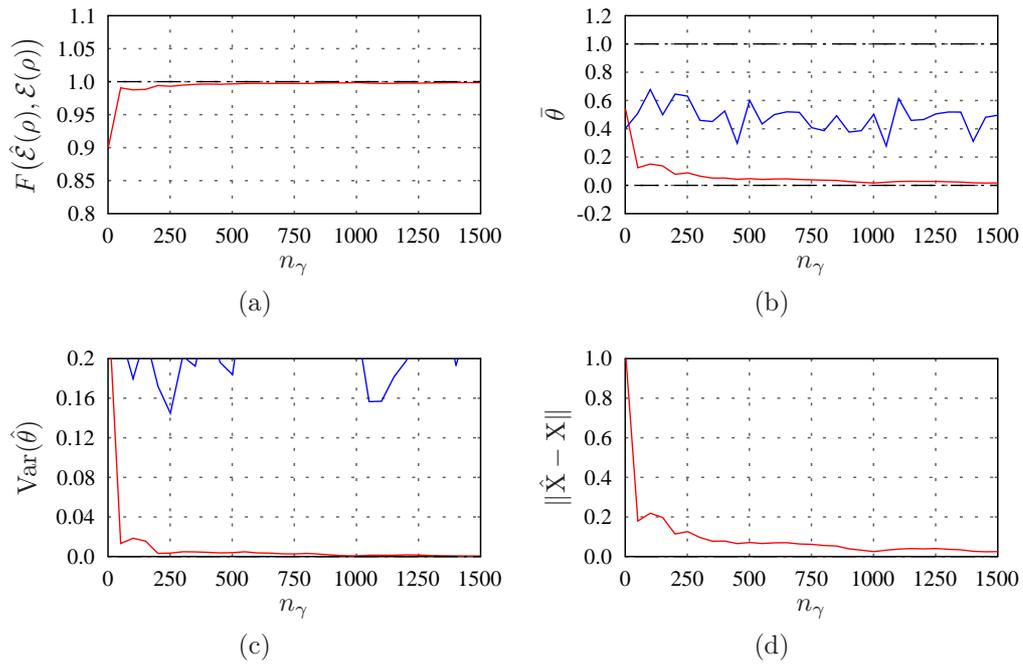

\begin{center}
\subfloat[]
{   
    \im{amp2-fid.eps}
    \label{fig:amp2-fid}
}
\subfloat[]
{   
    \im{amp2-mea.eps}
    \label{fig:amp2-mea}
}\\
\subfloat[]
{   
    \im{amp2-var.eps}
    \label{fig:amp2-var}
}
\subfloat[]
{   
    \im{amp2-hil.eps}
    \label{fig:amp2-hil}
}
\caption{The parameters are $\gamma=0$ and $p=1$. As $p$ drops out from the Choi matrix, it cannot be estimated. \subref{fig:amp1-fid} shows the fidelity of the estimated and ideal channel output, \subref{fig:amp1-mea} shows the expected value of the estimated parameters, \subref{fig:amp1-var} shows the variance of the estimated parameters, and \subref{fig:amp1-hil} shows the Hilbert--Schmidt norm of the estimation error.\label{fig:amp2}}
\end{center}
\end{figure}

\begin{figure}
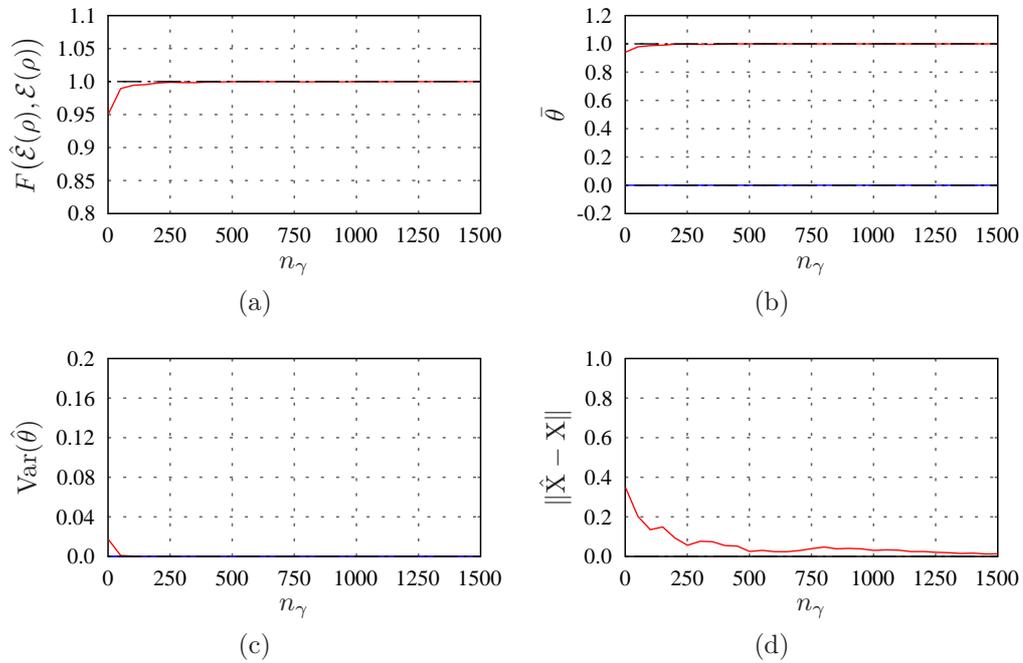

\begin{center}
\subfloat[]
{   
    \im{amp3-fid.eps}
    \label{fig:amp3-fid}
}
\subfloat[]
{   
    \im{amp3-mea.eps}
    \label{fig:amp3-mea}
}\\
\subfloat[]
{   
    \im{amp3-var.eps}
    \label{fig:amp3-var}
}
\subfloat[]
{   
    \im{amp3-hil.eps}
    \label{fig:amp3-hil}
}
\caption{The parameters are $\gamma=1$ and $p=0$. This is an estimation on the border of the parameter domain. \subref{fig:amp1-fid} shows the fidelity of the estimated and ideal channel output, \subref{fig:amp1-mea} shows the expected value of the estimated parameters, \subref{fig:amp1-var} shows the variance of the estimated parameters, and \subref{fig:amp1-hil} shows the Hilbert--Schmidt norm of the estimation error.\label{fig:amp3}}
\end{center}
\end{figure}

From the results. we can conclude that the estimated parameters are very accurate, even on the border of the parameter domain.


\subsection{T-representation of qubit Pauli channels}
\label{ssec:t.repr}
In the two dimensional case, we have the possibility to expand the quantum channels in the Pauli basis, thus giving the affine map of the channel, which takes Bloch vectors to Bloch vectors \cite{Petz-book}:
\[ 
\left[ \begin {array}{cccc}  1& 0\\\noalign{\medskip} t& T_3\end {array} \right] 
\] 
where $T_3$ can be assumed diagonal with elements $\alpha$, $\beta$, and $\gamma$ without loss of generality.

If we restrict ourselves to Pauli channels, i.e. we set $t=0$, then the Choi matrix of a quantum channel in this T-parametrization is
\[
\op{X}_{\cha{E}}=\left[ \begin {array}{cccc}  \frac{1+\gamma}{2}& 0& 0& \frac{\alpha+\beta}{2}\\\noalign{\medskip} 0& \frac{1-\gamma}{2}& \frac{\alpha-\beta}{2}& 0\\\noalign{\medskip} 0& \frac{\alpha-\beta}{2}& \frac{1-\gamma}{2}& 0\\\noalign{\medskip} \frac{\alpha+\beta}{2}& 0& 0& \frac{1+\gamma}{2}\end {array} \right]\ .
\] 
Here, the allowed values for the three parameters $\alpha$, $\beta$ and $\gamma$ for the channel to send density matrices to density matrices are 
\[
-1\leq\alpha,\beta,\gamma\leq 1\ , 
\]
and the positivity constraint is satisfied if and only if 
\[
|1\pm\gamma|\geq |\alpha\pm\beta|\ .
\]

Using \eqref{eq:complex.lmi}, the Choi matrix of this channel can be decomposed in the following way:
\begin{align*}
&H_0=\frac{1}{2}
{\scriptsize 
\left[ 
\begin {array}{cccc} 
 1 &  0  &  0  &  0\\ \noalign{\medskip}
 0 &  1  &  0  &  0\\ \noalign{\medskip}
 0 &  0  &  1  &  0\\ \noalign{\medskip}
 0 &  0  &  0  &  1
\end{array}
\right]
},
&&H_1=\frac{1}{2}
{\scriptsize 
\left[ 
\begin {array}{cccc} 
 0 &  0  &  0  &  1\\ \noalign{\medskip}
 0 &  0  &  1  &  0\\ \noalign{\medskip}
 0 &  1  &  0  &  0\\ \noalign{\medskip}
 1 &  0  &  0  &  0
\end{array}
\right]
},\\
&H_2=\frac{1}{2}
{\scriptsize 
\left[ 
\begin {array}{cccc} 
 0 &  0  &  0  &  1\\ \noalign{\medskip}
 0 &  0  & -1  &  0\\ \noalign{\medskip}
 0 & -1  &  0  &  0\\ \noalign{\medskip}
 1 &  0  &  0  &  0
\end{array}
\right]
},
&&H_3=\frac{1}{2}
{\scriptsize 
\left[ 
\begin {array}{cccc} 
 1 &  0  &  0  &  0\\ \noalign{\medskip}
 0 & -1  &  0  &  0\\ \noalign{\medskip}
 0 &  0  & -1  &  0\\ \noalign{\medskip}
 0 &  0  &  0  &  1
\end{array}
\right]
},
\end{align*}
and the optimization variables will be
\[
h_1=\alpha,\quad h_2=\beta,\quad h_3=\gamma\ .
\]

There are no convex relations in this case, so only the solution of \eqref{eq:ls.objective} is required. 

The fact, that the optimization variables $h_k$ are exactly the channel parameters themselves, makes the parameter extraction step \eqref{eq:paramextract.hs} also unnecessary for this channel. Thus, we can see that in this representation, the estimation of any two dimensional Pauli channel is a convex problem.   

The input state and the POVMs were the same in this experiment as in subsection \ref{ssec:gen.amp}, so the same three configurations were used. Note however, that in this example, the set of observables are not only tomographically complete, but also make it possible to estimate each channel parameter independently, resulting in a diagonal empirical covariance matrix. 

The resulting estimations, and characteristic quantities are plotted for three different sets of exact parameter values in Figures \ref{fig:trp1}-\ref{fig:trp3}. The chosen $(\alpha,\beta,\gamma)$ vectors were $(0.3,-0.1,0.1)$, $(0.5,0.3,0.2)$ and $(0.4,0.1,-0.5)$. These values were selected to represent both the interior and border of the parameter space. 
\begin{figure}
\begin{center}
\subfloat[]
{   
    \im{trp1-fid.eps}
    \label{fig:trp1-fid}
}
\subfloat[]
{   
    \im{trp1-mea.eps}
    \label{fig:trp1-mea}
}\\
\subfloat[]
{   
    \im{trp1-var.eps}
    \label{fig:trp1-var}
}
\subfloat[]
{   
    \im{trp1-hil.eps}
    \label{fig:trp1-hil}
}
\caption{The parameters are $\alpha=0.3$, $\beta=-0.1$ and $\gamma=0.1$. \subref{fig:trp1-fid} shows the fidelity of the estimated and ideal channel output, \subref{fig:trp1-mea} shows the expected value of the estimated parameters, \subref{fig:trp1-var} shows the variance of the estimated parameters, and \subref{fig:trp1-hil} shows the Hilbert--Schmidt norm of the estimation error.\label{fig:trp1}} 
\end{center}
\end{figure}

\begin{figure}
\begin{center}
\subfloat[]
{   
    \im{trp2-fid.eps}
    \label{fig:trp2-fid}
}
\subfloat[]
{   
    \im{trp2-mea.eps}
    \label{fig:trp2-mea}
}\\
\subfloat[]
{   
    \im{trp2-var.eps}
    \label{fig:trp2-var}
}
\subfloat[]
{   
    \im{trp2-hil.eps}
    \label{fig:trp2-hil}
}
\caption{The parameters are $\alpha=0.5$, $\beta=0.3$ and $\gamma=0.2$. \subref{fig:trp2-fid} shows the fidelity of the estimated and ideal channel output, \subref{fig:trp2-mea} shows the expected value of the estimated parameters, \subref{fig:trp2-var} shows the variance of the estimated parameters, and \subref{fig:trp2-hil} shows the Hilbert--Schmidt norm of the estimation error.\label{fig:trp2}} 
\end{center}
\end{figure}

\begin{figure}
\begin{center}
\subfloat[]
{   
    \im{trp3-fid.eps}
    \label{fig:trp3-fid}
}
\subfloat[]
{   
    \im{trp3-mea.eps}
    \label{fig:trp3-mea}
}\\
\subfloat[]
{   
    \im{trp3-var.eps}
    \label{fig:trp3-var}
}
\subfloat[]
{   
    \im{trp3-hil.eps}
    \label{fig:trp3-hil}
}
\caption{The parameters are $\alpha=0.4$, $\beta=0.1$ and $\gamma=-0.5$. \subref{fig:trp3-fid} shows the fidelity of the estimated and ideal channel output, \subref{fig:trp3-mea} shows the expected value of the estimated parameters, \subref{fig:trp3-var} shows the variance of the estimated parameters, and \subref{fig:trp3-hil} shows the Hilbert--Schmidt norm of the estimation error.\label{fig:trp3}} 
\end{center}
\end{figure}

As we can see, the parameter estimation in this convex case is very simple and accurate.


\subsection{The generalized Pauli channel in 3 dimension} 
\label{subsec:gen.pauli}
The generalized Pauli channel is discussed in \cite{Petz-2008}. It is a generalization of the Pauli channels for qubits into higher dimensions. Its definition for a general Hermitian operator $\op{A}$ taken from the operator algebra $\spa{L}(\spa{H})$: 
\[
\cha{E}(\op{A})=\left(1-\sum_{i=1}^{u}\lambda_i\right)\frac{\trace(\op{A})}{d}\hop{1}+\sum_{i=1}^{u}\lambda_iE_i(\op{A})\ ,
\]
where $d$ is the dimension of the input quantum system, and $u$ is the number of \emph{complementary subalgebras}\footnote{Two subalgebras $\mathcal{A}_1$ and $\mathcal{A}_2$ of $\spa{L}(\spa{H})$ are called complementary if the traceless subspaces of $\mathcal{A}_1$ and $\mathcal{A}_2$ are orthogonal with respect to the Hilbert--Schmidt inner product \cite{Petz-2007}.} of $\spa{L}(\spa{H})$ used in the channel construction. Let $\mathcal{A}_1,\dots,\mathcal{A}_u$ be pairwise complementary subalgebras, which linearly span the whole algebra $\spa{L}(\spa{H})$. Then $E_i$ is the orthogonal projection with respect to the Hilbert--Schmidt inner product onto $\mathcal{A}_i$.

Here, we consider the case when all of the complementary subalgebras are maximal Abelian. In this case the set of subalgebras correspond to a set of orthonormal bases called \emph{mutually unbiased bases} (MUB). Two orthonormal bases $\{\ket{\psi_{1,j}}\}$ and $\{\ket{\psi_{2,j}}\}$ are mutually unbiased if and only if the maximal Abelian subalgebras $\mathcal{A}_1$ and $\mathcal{A}_2$ containing operators diagonal in the bases $\{\ket{\psi_{1,j}}\}$ and $\{\ket{\psi_{2,j}}\}$ are complementary.\footnote{By another definition, two orthonormal bases $\{\ket{\psi_{1,j}}\}$ and $\{\ket{\psi_{2,j}}\}$ are mutually unbiased if for all $k,l=1,\dots,d$ they satisfy $|\braket{\psi_{1,k}}{\psi_{2,l}}|^2=\frac{1}{d}$. In $d$ dimensions, the maximal number of pairwise mutually unbiased bases is $d+1$. Currently, this maximal number of MUB are known to exist only in dimensions which are prime power \cite{Bandyopadhyay-2001}.} 

Let $\{\ket{\psi_{1,j}}\},\dots,\{\ket{\psi_{u,j}}\}$ be $u$ sets of bases which are pairwise mutually unbiased. Then the orthogonal projection $E_i$ can be constructed as
\[
E_i(\op{A})=\sum_{j=1}^d\bra{\psi_{i,j}}\op{A}\ket{\psi_{i,j}}\ketbra{\psi_{i,j}}{\psi_{i,j}}\ .
\]

The effect of the generalized Pauli channel is thus the following. It projects the input state onto each subalgebra $\mathcal{A}_i$, then scales the results with the corresponding channel parameter $\lambda_i$. Trivially, if $\lambda_i=0$ for all $i$, then the result is the completely mixed state $\frac{1}{d}\hop{1}$. In contrast, if $\lambda_i=1$ for all $i$, then we get back the input state as expected.

The channel is trace preserving by construction, and completely positive if and only if the parameters $\lambda_i$ satisfy
\begin{equation}
\label{eq:gen.pau.cp.cond}
1+d\lambda_i\geq\sum_j\lambda_j\geq\frac{-1}{d-1}\ ,
\end{equation}

A set of MUB can be constructed, for example, as follows \cite{Petz-2007}. Let $\ket{\varphi_0},\dots,\ket{\varphi_{d-1}}$ be a basis. Let $X$ and $Z$ be unitary operators such that 
\begin{gather*}
X\ket{\varphi_i}=\ket{\varphi_{i+1\!\!\!\mod (d-1)}}\ ,\\
Z\ket{\varphi_i}=\ee^{\ii k\frac{2\pi}{d}}\ket{\varphi_i}\ .
\end{gather*}
Then the unitaries $S_{j,k}\coloneqq Z^jX^k$ $(j,k=0,\dots,d-1)$ are pairwise orthogonal with respect to the Hilbert--Schmidt inner product. If we select the matrices $X$, $Z$, and also each one in the form $ZX^k$, $k=1,\dots,d-1$, then the eigenvectors of each of these unitaries form a set of MUB. Moreover, if $d$ is prime, then this procedure results in exactly $d+1$ MUB, which is the maximum number of MUB possible. 

In the present example, we choose $d=3$, thus the used MUB will be the bases obtained using the eigenvectors of the matrices $X$, $Z$, $ZX$, $ZX^2$. Then the Choi matrix of the channel will be the following:
\[
\op{X}_{\cha{E}}= \frac{1}{3}
{\scriptsize 
 \left[ \begin {array}{ccccccccc}  
 f_1  & 0   & 0   & 0   & f_3 & 0   & 0   & 0   & f_3   \\\noalign{\medskip} 
 0    & f_2 & 0   & 0   & 0   & f_4 &f_4^*& 0   & 0     \\\noalign{\medskip} 
 0    & 0   & f_2 &f_4^*& 0   & 0   & 0   & f_4 & 0     \\\noalign{\medskip} 
 0    & 0   & f_4 & f_2 & 0   & 0   &  0  &f_4^*& 0     \\\noalign{\medskip} 
 f_3  & 0   & 0   & 0   & f_1 & 0   & 0   & 0   & f_3   \\\noalign{\medskip} 
 0    &f_4^*& 0   & 0   & 0   & f_2 & f_4 & 0   & 0     \\\noalign{\medskip} 
 0    & f_4 & 0   & 0   & 0   &f_4^*& f_2 & 0   & 0     \\\noalign{\medskip} 
 0    & 0   &f_4^*& f_4 & 0   & 0   & 0   & f_2 & 0     \\\noalign{\medskip} 
 f_3  & 0   & 0   & 0   & f_3 & 0   & 0   & 0   & f_1   \end{array} 
 \right]\ ,}
\] 
where 
\begin{align*}
f_1&=1+ 2\lambda_2,\\
f_2&=1-\lambda_2,\\
f_3&=\lambda_1+\lambda_3+\lambda_4,\\
f_4&=\lambda_1-\frac{\lambda_3}{2}(1+\ii\sqrt{3})-\frac{\lambda_4}{2}(1-\ii\sqrt{3}),
\end{align*}
Note that the $n=2$ case using the same MUB selection method is exactly the same as the T-representation of Pauli channels in subsection \ref{ssec:t.repr}. 

The decomposition \eqref{eq:complex.lmi} of the Choi matrix can be calculated easily based on the previous examples, given that the optimization variables are exactly the channel parameters:
\[
h_i=\lambda_i,\quad i=1,\dots,4
\]
This, and the construction of the channel shows that the problem of parameter estimation is convex, and solvable using only \eqref{eq:ls.objective} in any such -- arbitrarily high -- dimension, in which the channel itself can be defined. 

The input state used in this experiment was the pure state
\[
\ket{\Psi}=\frac{1}{\sqrt{6}}\big(\ket{\psi_{1,1}}+\ket{\psi_{2,1}}+\ket{\psi_{3,1}}+\ket{\psi_{4,1}}\big)
\]
which is constructed from the first vector $\ket{\psi_{j,1}}$ of each basis of the MUB, on the analogy of the Bloch vector $\frac{1}{\sqrt{3}}[1,1,1]^\tp$ in the two dimensional case. Based on this analogy, this state is expected to be sufficient for the characterisation of the channel parameters. 

As we have discussed before, each channel parameter $\lambda_i$ affects the length of the projected input state $E_i(\rho)$ in the subalgebra $\mathcal{A}_i$ independently. The channel has no other effect. Thus, we can measure the effect of the channel by focusing only on the subalgebras, and estimate the parameters independently. For this purpose, any Hermitian operator belonging to the subalgebra $\mathcal{A}_i$ can be considered as observable, and can be used for the estimation of $\lambda_i$. The reason for this is that the effect of the channel is the same on the whole subalgebra, so it is enough to measure any direction inside $\mathcal{A}_i$ to get information on $\lambda_i$. Moreover, the independence of the measurements on different subalgebras result in a diagonal covariance matrix. 

For example, valid observables for each subalgebra can be constructed using the MUB in the following way: 
\[
\op{A}_i=\sum_{j=1}^dj\ketbra{\psi_{i,j}}{\psi_{i,j}}
\]
This way we get observables with $d$ different eigenvalues. Beyond that, the specific value of the eigenvalues are irrelevant, as we are interested only in the outcome probabilities.

\begin{figure}
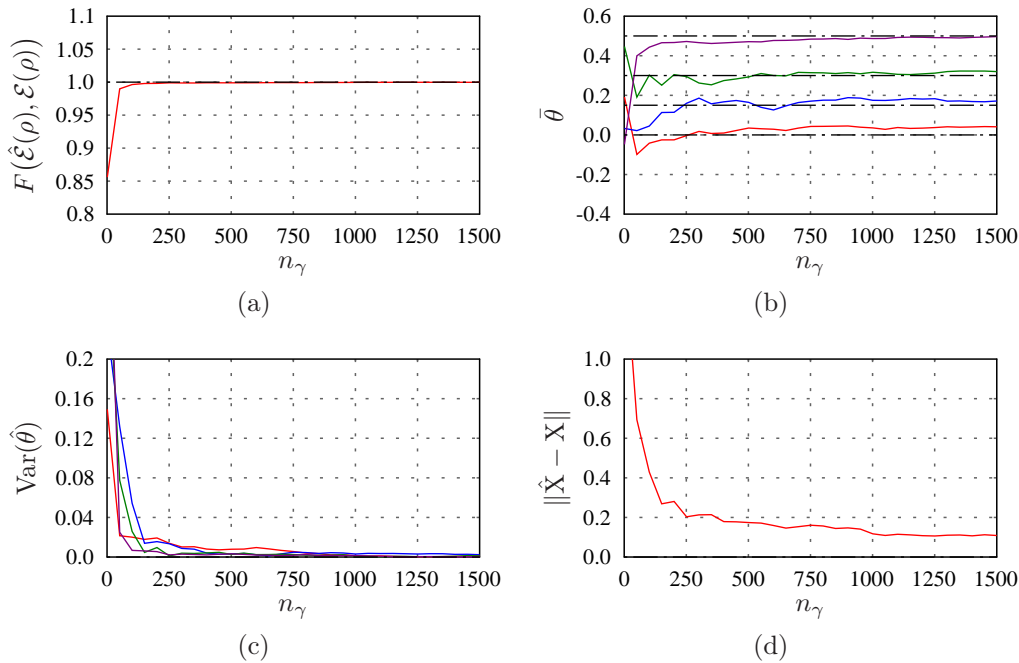

\begin{center}
\subfloat[]
{   
    \im{gpa1-fid.eps}
    \label{fig:gpa1-fid}
}
\subfloat[]
{   
    \im{gpa1-mea.eps}
    \label{fig:gpa1-mea}
}\\
\subfloat[]
{   
    \im{gpa1-var.eps}
    \label{fig:gpa1-var}
}
\subfloat[]
{   
    \im{gpa1-hil.eps}
    \label{fig:gpa1-hil}
}
\caption{$\lambda_1=0$, $\lambda_2=0.15$, $\lambda_3=0.3$ and $\lambda_4=0.5$. \subref{fig:gpa1-fid} shows the fidelity of the estimated and ideal channel output, \subref{fig:gpa1-mea} shows the expected value of the estimated parameters, \subref{fig:gpa1-var} shows the variance of the estimated parameters, and \subref{fig:gpa1-hil} shows the Hilbert--Schmidt norm of the estimation error.\label{fig:gpa1}}
\end{center}
\end{figure}

As it can be seen from the above, a total of four configurations were used in the experiments. The resulting estimations, and characteristic quantities are plotted for three different sets of exact parameter values in Figures \ref{fig:gpa1}-\ref{fig:gpa3}. The chosen $(\lambda_1,\lambda_2,\lambda_3,\lambda_4)$ vectors were $(0,0.15,0.3,0.5)$, $(0.1,0.3,0.4,0.5)$ and the third was $(-0.3,-0.2,-0.1,0.1)$. These values were selected to represent both the interior and border of the parameter space.

\begin{figure}
\begin{center}
\subfloat[]
{   
    \im{gpa2-fid.eps}
    \label{fig:gpa2-fid}
}
\subfloat[]
{   
    \im{gpa2-mea.eps}
    \label{fig:gpa2-mea}
}\\
\subfloat[]
{   
    \im{gpa2-var.eps}
    \label{fig:gpa2-var}
}
\subfloat[]
{   
    \im{gpa2-hil.eps}
    \label{fig:gpa2-hil}
}
\caption{$\lambda_1=0.1$, $\lambda_2=0.3$, $\lambda_3=0.4$ and $\lambda_4=0.5$. These saturate the first inequality in \eqref{eq:gen.pau.cp.cond}, so this is a border point of the parameter space. \subref{fig:gpa2-fid} shows the fidelity of the estimated and ideal channel output, \subref{fig:gpa2-mea} shows the expected value of the estimated parameters, \subref{fig:gpa2-var} shows the variance of the estimated parameters, and \subref{fig:gpa2-hil} shows the Hilbert--Schmidt norm of the estimation error.\label{fig:gpa2}}
\end{center}
\end{figure}

\begin{figure}
\begin{center}
\subfloat[]
{   
    \im{gpa3-fid.eps}
    \label{fig:gpa3-fid}
}
\subfloat[]
{   
    \im{gpa3-mea.eps}
    \label{fig:gpa3-mea}
}\\
\subfloat[]
{   
    \im{gpa3-var.eps}
    \label{fig:gpa3-var}
}
\subfloat[]
{   
    \im{gpa3-hil.eps}
    \label{fig:gpa3-hil}
}
\caption{$\lambda_1=-0.3$, $\lambda_2=-0.2$, $\lambda_3=-0.1$ and $\lambda_4=0.1$. These saturate the second inequality in \eqref{eq:gen.pau.cp.cond}, so this is again a border point. \subref{fig:gpa3-fid} shows the fidelity of the estimated and ideal channel output, \subref{fig:gpa3-mea} shows the expected value of the estimated parameters, \subref{fig:gpa3-var} shows the variance of the estimated parameters, and \subref{fig:gpa3-hil} shows the Hilbert--Schmidt norm of the estimation error.\label{fig:gpa3}}
\end{center}
\end{figure}

We can see that in the case of a qutrit channel with a high number of parameters, this method can provide very accurate parameter estimation, even on the border of the parameter domain. In some of the images, the Hilbert--Schmidt norm does not seem to decrease over the given range of the measurement number $n_\gamma$. Results from experiments performed with higher measurement numbers show that the reason of this is slower convergence. If measured this way, the desired estimation error is reached with a greater number of measurements in higher dimensions, due to the higher number of Choi matrix elements. 


\section{Conclusion} 
\label{sec:conclusion}
In this work, we introduced a method for the parameter estimation of quantum channel model families, based on convex optimization.

Using simulation case studies it has been shown, that the proposed method of affine decomposition of the Choi matrix can significantly increase the accuracy of the parameter estimation for the case of a known channel model structure. In addition, with performing auxiliary optimization problems, convex relations between optimization variables can also be exploited to improve the estimation. 

In some of the cases, particularly that of Pauli channels, this method results in a purely convex optimization problem, thus we can obtain a globally optimal estimation with relatively simple numerical algorithms. In the general case however, beside the convex part, the method may need a nonconvex optimization step as well.

\section*{Acknowledgement}
The authors would like to thank Professor Dénes Petz for helpful discussions, and Attila Magyar for the advices regarding the manuscript. 



\bibliographystyle{/home/scythia/_config/LaTeX.headers/diploma} 
\bibliography{/home/scythia/Munka/Doc.munka/Publications/cites-qpt}

\end{document}